Title:

Higher physical fitness levels are associated with less language decline in healthy ageing

Authors:

K. Segaert[1,2*], S.J.E. Lucas[3,2], C.V. Burley[3], P. Segaert[4], A. E. Milner[1], M. Ryan[5], L. Wheeldon[6,1]

Affiliations:

[1] School of Psychology, University of Birmingham, Birmingham, UK

[2] Centre for Human Brain Health, University of Birmingham, Birmingham, UK

[3] School of Sport, Exercise and Rehabilitation Sciences, University of Birmingham, Birmingham, UK

[4] Department of Mathematics, University of Leuven, Leuven, Belgium

[5] Cardiovascular Division, King's College London, London, UK

[6] Department of Foreign Languages and Translation, University of Agder, Kristiansand, Norway

* Address correspondence to:

Dr. Katrien Segaert

School of Psychology

University of Birmingham

Edgbaston

Birmingham

B15 2TT

UK

k.segaert@bham.ac.uk

**Abstract**

Healthy ageing is associated with decline in cognitive abilities such as language. Aerobic fitness has been shown to ameliorate decline in some cognitive domains, but the potential benefits for language have not been examined. In a cross-sectional sample, we investigated the relationship between aerobic fitness and tip-of-the-tongue states. These are among the most frequent cognitive failures in healthy older adults and occur when a speaker knows a word but is unable to produce it. We found that healthy older adults indeed experience more tip-of-the-tongue states than young adults. Importantly, higher aerobic fitness levels decrease the probability of experiencing tip-of-the-tongue states in healthy older adults. Fitness-related differences in word finding abilities are observed over and above effects of age. This is the first demonstration of a link between aerobic fitness and language functioning in healthy older adults.

Keywords: aerobic fitness, brain health, ageing, language, tip-of-the-tongue

**Introduction**

If the existing demographic trends continue, then in countries with high life expectancies such as the UK, Canada, the US and Japan, most children born after the year 2000 will live to become 100 years old[1]. In part motivated by the economic, healthcare and social challenges associated with this shift, there is an ever-growing interest in uncovering the antecedents of healthy ageing. In this paper we focus on cognitive changes, particularly changes in language abilities, in the healthy ageing population. Temporary cognitive lapses, such as not having a word come to mind when speaking, occur more frequently as we grow older. An interesting question therefore is whether lifestyle factors such as aerobic fitness are related to the occurrence of such 'senior moments'.

Maintaining good language skills is important for older adults. The experience of healthy ageing is only loosely related to one's numerical age. The way people move, feel, think, interact with and talk to others all co-determine the ageing experience. Focus groups have highlighted that maintaining social relations and independence are particularly instrumental for a positive experience of ageing[2], and good language abilities are crucial for achieving this. It is therefore also not surprising that when asked about age-related cognitive failures, older adults report that word finding difficulties are particularly irritating and embarrassing[3].

Word finding difficulties often surface as tip-of-the-tongue experiences. People in a tip-of-the-tongue state have a strong conviction that they know a word, but are unable to produce it. The frequency of tip-of-the-tongue states increases with age[4] and indeed tip-of-the-tongue states are documented to be among the most frequent cognitive failures in healthy older adults[5]. Older adults worry that tip-of-the-tongue states indicate serious memory problems[3]. However, this is a misconception: tip-of-the-tongue states are not associated with episodic memory loss[6]. In fact, older adults usually have a much larger vocabulary than younger adults[7]. Instead, focused experimental research has demonstrated that tip-of-the-tongue states are indicative of deficits in accessing phonology (i.e. sound form representations). Spoken word production is a two-stage process involving the retrieval of word meaning, followed by the associated phonology[8]. Tip-of-the-tongue states indicate a disruption in the process of transmission between meaning and phonology[4,9,10]. This process is essential for successful and fluent language production, and its disruption has very noticeable negative consequences for elderly speakers.

In the present research we investigated whether older adults' aerobic fitness levels are related to the incidence of age-related language failures such as tip-of-the-tongue states. Regular physical exercise is a lifestyle intervention approach that has received a considerable amount of attention in previous studies. As an intervention strategy, regular exercise is accessible, safe and effective[11]. Interventions as short as 6 weeks can result in a measurable increase in aerobic fitness[12]. Moreover, at least some modes of regular physical exercise are easily accessible, such as walking. Even when the primary mode of exercise that is taken up is walking, research has demonstrated short- and long-term cognitive benefits[13]. To date, no studies have investigated whether there is a relationship between aerobic fitness and language functioning. This stands in stark contrast to the amount of evidence of aerobic fitness benefits for other cognitive domains (e.g., cognitive control, executive functioning, visuo-spatial memory, learning and processing speed)[14,15].

Regular physical exercise and the resultant higher aerobic fitness is associated with reducing age-related decline in brain perfusion[16] and structural integrity[17]. Among others, structural integrity in frontal and temporal regions of the brain has been related to aerobic fitness[18,19]. This leads us to hypothesize that cognitive benefits of aerobic fitness may extend to the domain of language processing: language production is predominantly associated with functional activation in frontal and temporal regions in the brain[20]; word finding difficulties in particular are associated with functional activation[21] and structural atrophy[22] of the left insula. All this suggests that there may be aerobic fitness-related differences in word finding abilities; however to-date, no study has generated empirical evidence to support this hypothesis.

In the present study, we investigated the relationship between aerobic fitness and word finding abilities in a cross-sectional sample of healthy older adults. Word finding abilities were measured in a tip-of-the-tongue-eliciting experiment, in which the participants read definitions of words and were asked to produce the word. Aerobic fitness was quantified using a physiological measure of oxygen uptake from a graded exercise test. Using this gold-standard objective measurement of aerobic fitness, rather than a less reliable self-report physical activity measurement[23], we were able to accurately assess individual aerobic fitness levels. We hypothesized that there would be a positive relation between aerobic fitness levels and word finding abilities in our sample of healthy older adults, independent of age and vocabulary size. A demonstration of such a link could have far-reaching implications: the promise of potential ameliorating effects of regular physical exercise on an important and complex cognitive ability - language production - would add a crucial piece of knowledge to this growing and timely area of research.

# Results

*Older adults experience more tip-of-the-tongue occurrences and have less access to phonological information, compared to young participants*

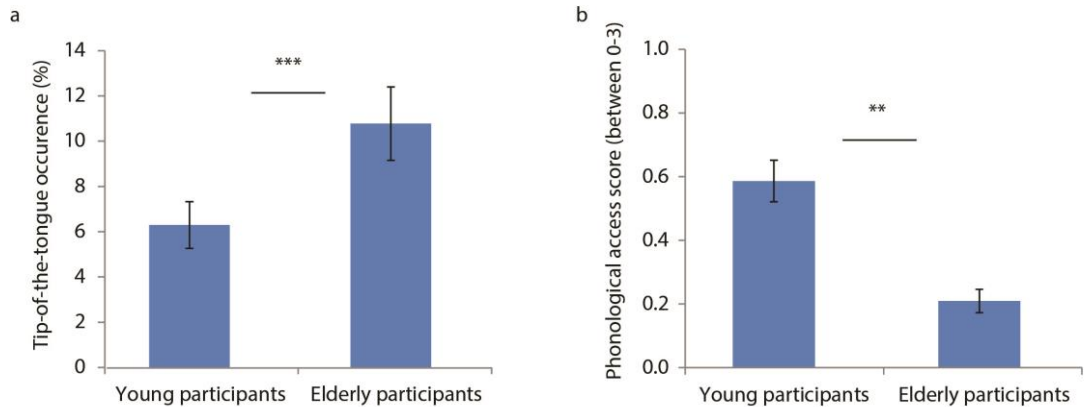

**Figure 1. Age-related differences in word finding abilities.** (a) The percentage of tip-of-the-tongue states experienced by the older adults was higher than for young participants. (b) When experiencing tip-of-the-tongue states, older adults had access to significantly less correct phonological information than young participants. Error bars represent standard error of the mean.

Figure 1 summarizes the age-related differences in healthy older adults compared to young controls with regard to word finding abilities.

We first investigated the effect of age on tip-of-the-tongue occurrence (Table 1A). In line with our predictions, we found that older adults experienced more tip-of-the-tongue occurrences than young participants ($p<0.001$) (Figure 1a). Both young and older adults experienced fewer tip-of-the-tongue occurrences the shorter the target words (as measured by the number of phonemes) ($p<0.001$) and the larger their vocabulary size ($p<0.001$).

Next we looked at the age-related differences in the access of correct phonological information about the words (Table 1B). When experiencing a tip-of-the-tongue state, older adults had less access to correct phonological information than young participants ($p<0.002$) (Figure 1b).

We found also that older adults had a significantly larger vocabulary than young participants ($t(53)=5.01$, $p<0.001$), which is in line with previous observations.

**Table 1. Summary of the mixed effects regression models predicting tip-of-the-tongue occurrence and phonological access in young versus older adults**

*A. Mixed effects logistic regression model predicting tip-of-the-tongue occurrence*

|  | Coefficient | SE | Wald z | p value | |
|---|---|---|---|---|---|
| Intercept | -3.51 | 0.24 | -14.52 | <0.001 | *** |
| Age group | -0.75 | 0.14 | -5.24 | <0.001 | *** |
| No. phonemes of target word | 0.30 | 0.08 | 3.60 | <0.001 | *** |
| Vocabulary size | -0.06 | 0.01 | -4.30 | <0.001 | *** |

Note: N = 3300, AIC = 1699.5, log-likelihood = -839.8
This model includes a random intercept for items and participants, a random slope for Vocabulary Size for items and a random slope for the Number of Phonemes of the Target Word for participants. Multicollinearity was low (all VIF < 1.4)

*B. Mixed effects linear regression model predicting phonological access scores*

|  | Coefficient | SE | df | t value | p value | |
|---|---|---|---|---|---|---|
| Intercept | 0.51 | 0.08 | 47.31 | 6.38 | <0.001 | *** |
| Age group | 0.24 | 0.07 | 38.01 | 3.32 | <0.002 | ** |

Note: N = 272, AIC = 568.58.
This model includes a random intercept for items and participants
*** < .001   ** < .01   * < .05

*Healthy older adults' aerobic fitness levels are related to the incidence of tip-of-the-tongue states*

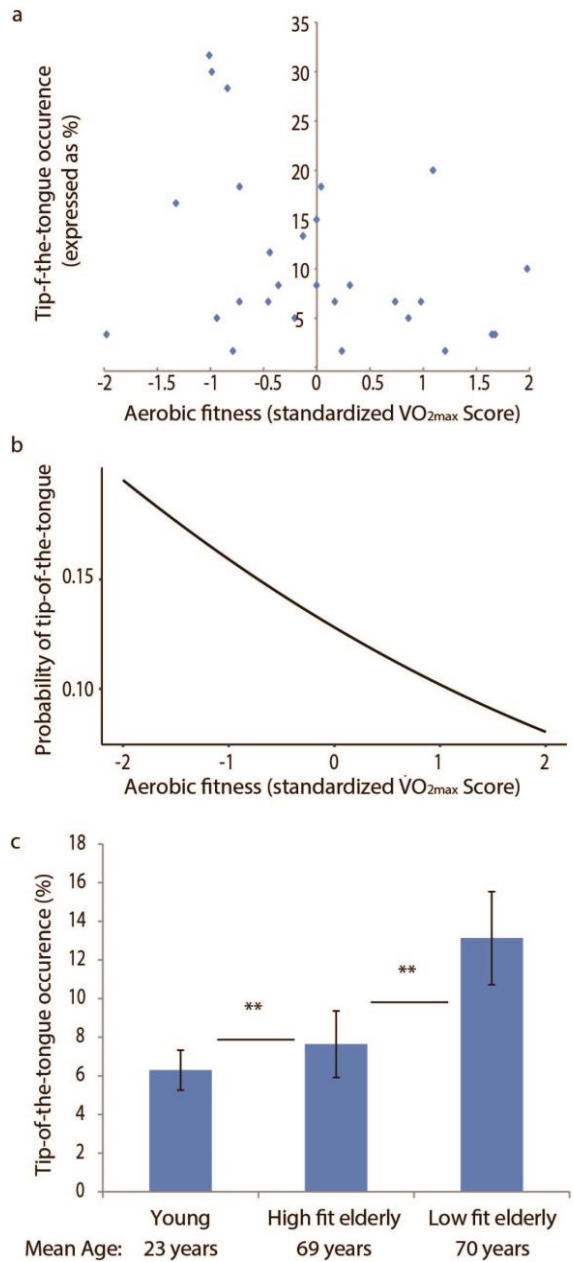

**Figure 2. Aerobic fitness-related differences in word finding abilities.** (a) Depicted for older adults is the tip-of-the-tongue occurrence as a function of standardized aerobic fitness scores. (b) There was a significant influence of the standardized aerobic fitness scores on the probability of experiencing a tip-of-the-tongue state. Depicted is the mean effect across participants, keeping all other variables in the model (see table 2) constant. The higher the aerobic fitness score, the lower the probability of experiencing a tip-of-the-tongue state. (c) When the older adults were divided into groups of high- and low-fitness (high- and low-fit as determined by a median split, groups were matched for age, education level and vocabulary size), we found that low-fit older adults had a higher incidence of tip-of-the-tongue states than high-fit older adults. High-fit older adults in turn had a higher incidence than the young participants. Error bars represent standard error of the mean.

Next we investigated the relationship between aerobic fitness and word finding abilities in healthy older adults.

First, we tested whether aerobic fitness scores were predictive of the probability of experiencing a tip-of-the-tongue state. Our objective physiological marker of aerobic fitness (standardized $\dot{V}O_{2max}$ scores) indeed significantly predicted tip-of-the-tongue occurrences (p=0.024). The more aerobically fit the older adults were, the less likely they were to experience a tip-of-the-tongue state. The coefficient of each predictor in the model is informative with regard to the size of its effect: an increase in aerobic fitness ($\dot{V}O_{2max}$) by one standard deviation is related to a 28 percent decrease in the probability of experiencing a tip-of-the tongue experience on the logit scale. To explain this further, consider for example a hypothetical situation where an elderly person has a 10 percent chance of experiencing a tip-of-the tongue state. Assuming that the number of phonemes of the target word, age and vocabulary size remains constant, an increase in aerobic fitness by one standard deviation is associated with a reduction in the chance of experiencing a tip-of-the-tongue state to 7.75 percent. Figure 2b depicts the influence of aerobic fitness scores on the probability of experiencing a tip-of-the-tongue state, keeping all other variables in the model constant (the figure shows the mean effect across participants).

The relationship between aerobic fitness and tip-of-the-tongue occurrence was observed over and above the effects of age and vocabulary size. Effects of age and vocabulary size were also accounted for in the model. Tip-of-the-tongue occurrences increased with participant age ($p=0.027$). Also, the larger the vocabulary size of the older adults, the less likely they were to have tip-of-the-tongue experiences ($p<0.001$). Older adults experience more tip-of-the-tongue occurrences for longer words, as measured by the number of phonemes of the target word ($p<0.001$). Education level did not contribute toward predicting tip-of-the-tongue states – a model including this effect was not a better fit than a model excluding this effect ($\chi^2_1 = 0.88$, $p>.35$). This may be because there was relatively little variation in education level among our participants, with only a few who had not received formal education at university level. Also, a model including sex was not a better fit than a model excluding this effect ($\chi^2_1 = 0.01$, $p>.95$).

The result of the mixed effects logistic regression model of tip-of-the-tongue occurrence in the group of older adults is summarized in table 2.

A mixed effects linear regression model on phonological access scores (including a fixed effect for aerobic fitness, a random intercept and slope for aerobic fitness score for items and a random intercept for participants), revealed that aerobic fitness scores were not predictive of phonological access ($\beta = -0.02$, SE=0.10, df=17.27, t=-0.15, $p>.8$).

Lastly, we performed a median split on the standardized aerobic fitness scores to create a high-fit older adults and a low-fit older adults group. Comparison of group means revealed that the two groups did not differ in age (no assumption of equal variance: $t_{(21.17)}=-1.05, p>0.3$), vocabulary size ($t_{(26)}=.20, p>0.8$) or education level ($t_{(26)}=-.60, p>0.5$). Despite the fact that high-fit and low-fit older adults are thus matched on age, education level and vocabulary size, the tip-of-the-tongue occurrence of high-fit older adults was lower than that of low-fit older adults ($\beta = 0.64$, SE=0.29, Wald z=2.22, $p<.03$), although still higher than that of young adults ($\beta = -1.12$, SE=0.33, Wald z=3.43, $p<0.001$) (model including fixed effects for the number of phonemes of target word and vocabulary size, random intercepts for items and participants) (Figure 2c).

In sum, the data show a relationship between aerobic fitness and word finding abilities in a group of healthy older adults.

**Table 2. Summary of the mixed effects logistic regression model predicting tip-of-the-tongue occurrence in the older adults**

|  | Coefficient | SE | Wald z | p value |  |
|---|---|---|---|---|---|
| Intercept | -3.17 | 0.26 | -12.04 | <0.001 | *** |
| No. phonemes of target word | 0.32 | 0.09 | 3.68 | <0.001 | *** |
| Vocabulary size | -0.06 | 0.01 | -5.48 | <0.001 | *** |
| Age | 0.05 | 0.02 | 2.22 | 0.027 | * |
| Aerobic fitness ($\dot{V}O_{2max}$ score) | -0.28 | 0.12 | -2.25 | 0.024 | * |

Note: N = 1680, AIC = 998.7, log-likelihood = -492.3, *** < .001   ** < .01   * < .025
This model included a random intercept for items and participants. Multicollinearity was low (all VIF < 1.2).

## *Power of detecting an aerobic fitness effect on tip-of-the-tongue occurrence in healthy older adults*

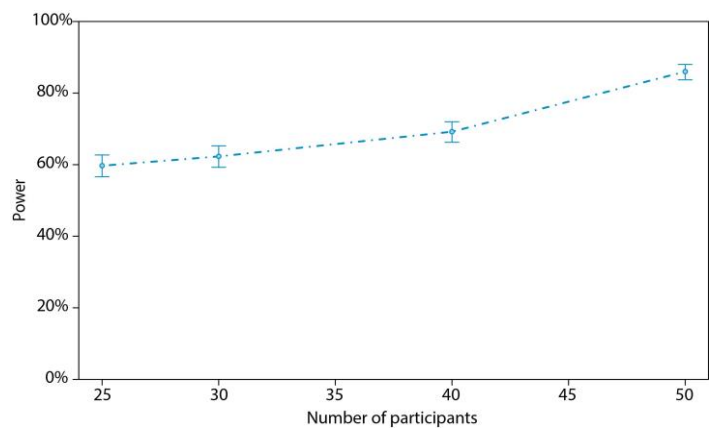

**Figure 3. Power of the aerobic fitness effect on tip-of-the-tongue occurrence in function of the number of participants.** Power is estimated as the chance of detecting a non-zero effect of aerobic fitness on the probability of experiencing a tip-of-the-tongue state in a simulation study with 1000 simulation runs. The figure indicates that already a small sample size leads to a high power of the study. Error bars represent the 95% confidence interval for the proportion of simulation runs in which the null-hypothesis was rejected.

The chance of a Type-I error (i.e. incorrectly rejecting the null-hypothesis) for all above described effects is fixed to 5%, since the alpha level for the study was 0.05. With the aim to provide information on Type-II error in possible future studies, we investigated the power of detecting a non-zero effect of aerobic fitness on the probability of experiencing a tip-of-the-tongue state by means of a simulation study. Under the above fitted model, we simulated new independent data sets with a given number of participants using the R-package simr[24]. The power then signals the percentage of those simulated data sets for which the hypothesis of a zero effect is rejected. Figure 3 shows the obtained power for 1000 simulation runs in function of the number of participants included in the study. For a sample of 25 and 30 participants, a respective power of 59.70% (confidence interval [56.59, 62.76]) and 62.30% (confidence interval [59.21, 65.31]) is obtained. Note that our current study uses N = 28 participants. Figure 3 can be used to inform sample size estimation and power analyses in future studies.

**Discussion**

We demonstrate for the first time that there is a relationship between language production abilities and aerobic fitness in healthy older adults. The data reveal that healthy older adults have more tip-of-the tongue occurrences, and when they do, have less access to phonological information about the target word in comparison to the young control group. However, most importantly, older adults' aerobic fitness levels are related to the incidence of age-related language failures such as tip-of-the-tongue states. Specifically, there is a relation between tip-of-the-tongue occurrence and aerobic fitness over and above the influence of age and vocabulary size. The higher the older adults' aerobic fitness level, the lower the probability of experiencing a tip-of-the-tongue state.

There is an increasing interest in lifestyle factors that could ameliorate age-related decline. A growing number of studies have demonstrated a relationship between regular physical exercise and different domains of cognition[14], but the present study is to the authors' knowledge the first to investigate the link between fitness and language processing. Language is a crucial aspect of cognition, necessary for maintaining independence, communication and social interaction in older age. Clearly, language processing is related to general cognitive functions such as processing speed, executive functions, working memory and declarative memory. However, language functioning cannot be reduced to functioning in these non-linguistics domains and the relationship between language decline and the decline in these non-linguistics cognitive functions is not yet fully understood. It is therefore noteworthy that we are able to show, for the first time, that the benefits of aerobic fitness extend to the domain of language.

Our finding that older adults experience more tip-of-the-tongue states, and have less access to correct phonological information when in a tip-of-the-tongue state is in line with previous observations[4,9,10]. Previous research on the impact of ageing on language processing has identified that older adults have reduced word finding abilities, and also have syntactic processing difficulties when speaking. Consequently, older adults choose to produce less complex syntactic structures and make more grammatical errors, e.g. [25]. One interesting implication from the findings of the current study is that future research could explore whether higher aerobic fitness levels also relate to age-related decline in this aspect of language processing.

Considerable progress has been made in elucidating the changes in brain function and structure that underlie the cognitive benefits associated with higher levels of physical activity, which ultimately manifest in increases in aerobic fitness. Firstly, greater aerobic fitness is associated with vascular benefits for the ageing brain. Effective regulation of brain blood flow, including the effectiveness of blood vessels to respond to changes in the concentration of carbon dioxide ($CO_2$), is vital for optimal brain function[11,26]. Cerebral blood flow and cerebrovascular reactivity to $CO_2$ decrease with age, but higher aerobic fitness levels are associated with less decline in these measures of brain function[16]. Moreover, vascular function has been shown to improve following an exercise training intervention aimed at increasing aerobic fitness[27]. Several studies have demonstrated a relationship between vascular brain functioning and various aspects of cognitive performance, including cognitive control[28] and speed[29]. Secondly, aerobic fitness is related to neural effects in the ageing brain. Healthy ageing leads to reductions in grey matter volume and altered integrity of white matter tracts[30], however higher aerobic fitness levels have been demonstrated to act as a protector for age-related decline in the brain's structural integrity[17]. Higher aerobic fitness is associated with greater grey matter volume in frontal and hippocampal regions[31]. Moreover exercise training in sedentary older adults is demonstrated to have widespread effects, sparing tissue in frontal and temporal regions[18], and has even been shown to increase volume and thus reverse age-related grey matter loss in the hippocampus[32]. Higher aerobic fitness is also predictive of white matter integrity, in the corpus callosum[33] and frontal and temporal lobes[19]. These beneficial effects on the brain's structural integrity have been linked to cognitive improvements in memory functioning[19,32]. A few studies have investigated the influence of aerobic fitness on functional neural activation, finding a relationship between activation in task-relevant brain regions and fitness during a processing speed task[13] and during attention processing[34]. Despite the clear progress that has thus been made in elucidating the benefits of aerobic fitness for brain structure and function, the complex relationship between brain structure, brain function and cognition in this context still remains poorly understood.

More research is needed therefore to study how the relationship between brain structure, brain function and language abilities is changed by aerobic fitness and regular exercise. Previous research on healthy older adults has demonstrated that tip-of-the-tongue states are associated with grey matter atrophy in the left insula[18] and functional activation changes in linguistic (phonological processing: left insula[19]) as well as extra-linguistic brain networks (cognitive control: anterior cingulate and lateral prefrontal cortex[20,21]). Future research could focus therefore on whether structural and functional activation changes in these brain regions underlie fitness-related differences in word finding abilities.

In the discussion of the literature above, the findings from cross-sectional research and intervention-type research led to converging conclusions with regard the influence of aerobic fitness on cognition, brain structure and brain function. It must be emphasized that the present research is cross-sectional in nature and thus no causal conclusions can be drawn. Potential confounds in our study are the amount of social interaction and language input that older adults are exposed to. These may correlate with greater mobility and higher aerobic fitness levels. Future research will have to determine whether an exercise intervention can successfully increase language abilities. We suggest that such an intervention study would have an active control group that would be exposed to an enriched environment with increased social interaction, cognitive stimulation and language input – similarly to the aerobic exercise group. Such an intervention study would thus be able to determine whether aerobic training *per se* successfully improves language abilities.

In summary, in this paper we find an important relationship between age-related decline in language production abilities and aerobic fitness. Word production is an essential step in successful and fluent language production, and the disruption of this process in healthy ageing is detrimental for a positive ageing experience. The present results suggest that higher aerobic fitness levels are associated with better word production skills in healthy older adults, and thus further support the promotion of increased physical activity for healthy ageing and optimal brain function across the life span.

## Methods

### Participants

53 older adults (34 women, mean age: 70.9 years, SD: 4.4; 19 men, mean age 69.4, SD: 4.9) volunteered to participate in the study. These participants all completed health screening to minimise the risk of an adverse event occurring during the exercise test. Screening information was reviewed by a cardiologist, resulting in 25 older adults being excluded from participating in aerobic fitness testing (for details on exclusion criteria, see below). Consequently, 28 older adults (20 women, mean age: 70.3 years, SD: 4.4; 8 men, mean age: 67.6 years, SD: 5.0) completed the aerobic fitness test and the language experiment. The average education level of the 28 older adults was 16.4 years (SD: 3.2) of formal education, which in the UK starts at 4 years old. The average height was 165.4 cm (SD: 10.5) and the average weight was 66.9 kg (SD: 10.7). For all but 3 older adults we obtained a MOCA score (Montreal Cognitive Assessment) and they scored 26 or higher, which is considered normal.

To provide a baseline against which to compare the older adults' language abilities, 27 young participants (19 women, mean age: 23.4, SD:3.9; 8 men, mean age: 22.9, SD: 2.5) completed the language experiment. The young participants did not complete aerobic fitness testing. All young participants were currently enrolled as university students with the University of Birmingham.

All participants gave informed consent and were monetarily compensated for participation. All were non-bilingual native British English speakers with no speech or language disorders and no dyslexia. The research was conducted at the University of Birmingham. The research had full ethical approval (UoB ERN_16-0230) and all experimentation was performed in accordance with the relevant guidelines and regulations.

### Electrocardiogram (ECG) and general health screening

53 older adults underwent a screening procedure prior to the exercise testing. The evaluation consisted of a general health questionnaire, a resting 12-lead electrocardiogram (ECG) assessment and a resting blood pressure measurement. This information was then reviewed by a cardiologist (MR). Participants who revealed a contraindication to non-medically supervised exercise testing in the general health questionnaire (N=5; e.g., heart condition, family history of heart attack, asthma, prevention medication for stroke), had high resting blood pressure (N=3; systolic >160, diastolic >90), or showed ECG abnormalities (N=12; e.g., ST depression, abnormal QRS axis, abnormalities of cardiac rhythm) were excluded from the aerobic fitness testing (and referred on to their GP).

Furthermore, for 3 older adults who passed the screening protocol, exercise testing on the ergometer had to be terminated before a fitness score was obtained because the participants experienced knee pain. Two older adults choose to withdraw participation post-screening. As a result, fitness scores were obtained for 28 older adults.

In the retained sample of 28 older adults who completed fitness testing, 4 older adults were taking anti-hypertension medication, 3 older adults were on cholesterol lowering medication, and 1 older adult was taking both types of medication.

**Aerobic fitness testing**

After screening and inclusion into the study, participants completed a graded sub-maximal aerobic fitness test on a cycle ergometer to estimate maximal oxygen consumption ($\dot{V}O_{2max}$). The sub-maximal fitness test was based on the Åstrand-Rhyming Cycle Ergometer Test, which has been shown to provide a reliable and valid estimate of $\dot{V}O_{2max}$ [35]. Submaximal estimation of maximal aerobic power is a standard procedure for measurement of fitness in sedentary older adults and clinical populations.

For this test participants were asked to cycle on an electromagnetically braked cycle ergometer at 60-70 rpm (rotations per minute). The initial workload began at 35 Watts and then depending on the participant's sex, body mass and habitual physical activity levels, workload increased by 20 to 35 Watt increments every three minutes. This continued until heart rate reached 80% of the participant's estimated maximum heart rate (i.e. 220 minus the participant's age), unless the participant was unable to maintain over 50 rpm or until the participant reached volitional exhaustion. Respiratory gases and volume were collected for measurement of the rate of oxygen consumption ($\dot{V}O_2$). Maximal $\dot{V}O_2$ was then estimated from the relationship between oxygen uptake and heart rate at multiple measurements. The resulting regression equation predicted participant's $\dot{V}O_{2max}$ (as per the standard procedure: Guiney, et al. [28], Siconolfi, et al. [35]). Prior to this test, participants were asked to abstain from heavy physical exercise and alcohol for 24 hours. They were also instructed not to consume food for 2 hours prior to reporting to the laboratory.

For female participants, the average predicted $\dot{V}O_{2max}$ score was 23.32 (SD = 7.04) with values ranging from 9.4 to 35.1. For male participants, the average predicted $\dot{V}O_{2max}$ score was 31.16 (SD = 6.55) with values ranging from 24.7 to 44.1. It is a standard finding that males have higher $\dot{V}O_{2max}$ scores than females [36]. In general, males have larger body mass (including lung size and cardiovascular capacity) than females, so direct comparison of the raw $\dot{V}O_{2max}$ score for a male and a female is not valid [37]. Scores therefore must be normed or standardized. We calculated z-scores within each sex group for the purpose of relating $\dot{V}O_{2max}$ scores to tip-of-the-tongue occurrence. Using the standardized score allows male and female participants to be viewed on one and the same dimension with regard to $\dot{V}O_{2max}$ scores.

**Tip-of-the-tongue experiment**

Participants completed a definition filling task: a definition appeared on screen, and participants were asked to indicate whether they knew the word (No/Yes, produce the word) or had a tip-of-the-tongue experience.

The definition materials consisted of 20 definitions of low frequency words (adapted from Jones [38]), 20 questions about people famous in the UK, such as authors, politicians and actors (some adapted from [39]), and 20 definitions of easy words – see Table 3 for examples. Each participant received the 60 definitions in a random order.

The sequence of events on each trial was as follows. A warning signal was displayed for 500 ms after which a definition appeared centred on the screen. The definition remained on screen until the participant responded as follows: they knew the word (button press 'Yes', and then said the word out loud), did not know the word (button press 'No'), or had a tip-of-the-tongue experience (button press 'ToT'). In the instructions to the participants we defined a tip-of-the-tongue experience as: *"Usually we are sure if we know or don't know a word. However, sometimes we feel sure we know a word but are unable to think of it. This is known as a 'tip-of-the-tongue' experience"*.

If participants indicated they experienced a tip-of-the-tongue state, they were asked to provide three pieces of information about its sound structure in response to prompts on the screen which asked them to: 1) guess the initial letter or sound; 2) guess the final letter or sound, and 3) guess the number of syllables. Finally, in order to determine if they were correct in thinking that they knew the target word, participants were asked to select it from a list of four words that were displayed on the screen (the correct answer and three foils – see Table 3 for examples) or to indicate that the word they were thinking of was not in the list.

**Table 3: Examples of definitions, target words and foils for multiple-choice questions if participants indicated to have experienced a tip-of-the-tongue.**

| Definitions | Correct answer | Foils |
|---|---|---|
| 1. A young goose | Gosling | cygnet, leveret, gelding |
| 2. Able to read and write | Literate | laconic, loquacious, urbane |
| 3. What is the original last name of the boxer who became known as Mohammed Ali? | Clay | Grey, Reid, Grant |
| 4. What was Princess Diana's maiden name? | Spencer | Ogilvy, Lawrence, Philips |
| 5. Ancient tomb for Egyptian kings | Pyramid | crypt, grave, sphynx |
| 6. A fruit of the oak tree eaten by squirrels | Acorn | nut, pit, seed |

**Data analyses**

We analysed the data using mixed effects models, which are an extension of classical linear regression models. Mixed effects models are the most suitable models to analyse the present dataset because they can account for the fact that there are repeated observations for both items and participants. We modelled tip-of-the-tongue occurrence using mixed effects logistic regression[40,41] in R[42].

For a categorical outcome variable such as tip-of-the-tongue occurrence, a logistic regression is much more suited than an ANOVA to model the data[40]. Using ANOVA models when the dependent variable is categorical (e.g., yes/no, counts, percentages) can lead to spurious significance values[40,43,44]. In such instances, regression methods are thus preferred[45]. However, ordinary regression analysis ignores correlation of observations within clusters and treats within cluster observations the same as between cluster observations producing invalid standard errors of the fitted coefficients[46]. Any subsequent analysis based on these standard errors (e.g., hypothesis test) is therefore invalid. The use of mixed effect models allows accounting for the fact that there are repeated observations for both items and participants[40,41] and therefore used frequently in psycholinguistic literature.

In addition to modelling tip-of-the-tongue occurrence, we also fitted a model for phonological access scores. We calculated a phonological access score for each trial on which the participant reported a tip-of-the-tongue: 1 point for listing the correct initial sound, 1 point for the correct final sound and 1 point for the correct number of syllables (resulting in a score between 0 and 3). Phonological access scores were modelled using mixed effects linear regression[41] in R[42], again to account for the fact that there are repeated observations for both items and participants.

The regression models for tip-of-the-tongue occurrence and phonological access scores were based on the following predictors: Number of Phonemes of the Target Word, Number of Syllables of the Target Word, Vocabulary Size (% of items named correctly), Education Level (years of formal education), and Age Group (young vs. older adults) / Age (in years). When modelling tip-of-tongue occurrence in the older adults group, we also included a predictor with the standardized $\dot{V}O_{2max}$ scores (see above).

Continuous variables were centered. Group was deviation coded. We started with including a maximal random effect structure as justified by the design and in the case of non-convergence we simplified the random effects structure until convergence was reached. The random effect associated with the smallest variance is dropped and this is done progressively until convergence is reached[41]. During the process of model comparison, we started with a model including all fixed effects and then simplified the model using model comparison for fixed effects in stepwise fashion until a model was reached with the lowest AIC value (Akaike information criterion).

Main models are summarized in tables; coefficient estimates are included in the text only when a full summary is not included in the tables.

**Sample size estimation and power analysis**

No previous study has investigated the effects of aerobic fitness on any aspect of language functioning; previous studies have so far reported effects only of aerobic fitness in other cognitive domains: cognitive control, executive functioning, visuo-spatial memory, learning and processing speed (Colcombe & Kramer, 2003). Thus there is no direct evidence on which we could postulate a hypothesized effect size prior to conducting the present study. Rather than postulating a hypothesized effect size based on indirect evidence, we performed a power analysis simulation study which can be a basis for future work. Indeed, any result of a power calculation depends entirely on the size of the hypothesized effect, for which it is impossible to obtain an accurate estimate until direct evidence from a first study is available. The current research may then serve as a benchmark for future studies.

We investigated the power of detecting a non-zero effect of aerobic fitness on the probability of experiencing a tip-of-the-tongue state by means of a simulation study under the statistical model describing the present data. As such, we can, based on informative evidence, investigate the Type II error which depends on both the effect size and the sample size. The simulation study allows us to generate data independent of the data described in the current work. Moreover, simulation studies are an effective way of obtaining a power estimate for complex models[47-49] and the approach has been used in the field of psychology in recent years[50-52]. The results of our simulation will serve as in important indicator for postulating sample size in future studies. Further details are described in the results section.

**Acknowledgments**

This research was funded by a Wellcome Trust ISSF Award to KS, SL and LW. We would like to thank Prof. Janet Lord for helpful feedback during a workshop on Ageing and Health in Sao Paulo when we were conceiving this research. We would also like to thank Denise Clissett for help with participant recruitment, Beth Skinner for her assistance collecting the ECG data, Dr. Evelien Heyselaar for much appreciated advice on the data analyses, and Prof. Ole Jensen for editing the manuscript.

**Author contributions**

K.S, S.J.E.L. and L.W. designed the study. C.V.B. collected the ECG and aerobic fitness data. A.E.M. collected the language data. M.R. was responsible for the medical screening. P.S. performed power simulations. K.S. analysed the data and wrote the manuscript. All authors edited the manuscript.

**Competing interests**

The authors declare no competing interests.